\documentclass{article}
\usepackage{spconf,graphicx,amsmath,amssymb,amsfonts}
\usepackage[OT1]{fontenc} 
\usepackage{cite}
\usepackage{algorithm}
\usepackage{algorithmic}
\usepackage{multirow}
\usepackage{url}

\title{Consistency-aware multi-channel speech enhancement \\ using deep neural networks}
\name{Yoshiki Masuyama$^\dagger$\thanks{This work was done while Yoshiki Masuyama was in LINE Corporation.}, Masahito Togami$^\ddag$, Tatsuya Komatsu$^\ddag$}
\address{
$^\dagger${\fontsize{11pt}{0pt}\selectfont Department of Intermedia Art and Science, Waseda University, Tokyo, Japan} \\
$^\ddag${\fontsize{11pt}{0pt}\selectfont LINE Corporation, Tokyo, Japan}
}
\begin{document}
\ninept
\maketitle

\begin{abstract}
This paper proposes a deep neural network (DNN)--based multi-channel speech enhancement system in which a DNN is trained to maximize the quality of the enhanced time-domain signal.
DNN-based multi-channel speech enhancement is often conducted in the time-frequency (T-F) domain because spatial filtering can be efficiently implemented in the T-F domain.
In such a case, ordinary objective functions are computed on the estimated T-F mask or spectrogram.
However, the estimated spectrogram is often inconsistent, and its amplitude and phase may change when the spectrogram is converted back to the time-domain.
That is, the objective function does not evaluate the enhanced time-domain signal properly.
To address this problem, we propose to use an objective function defined on the reconstructed time-domain signal.
Specifically, speech enhancement is conducted by multi-channel Wiener filtering in the T-F domain, and its result is converted back to the time-domain.
We propose two objective functions computed on the reconstructed signal where the first one is defined in the time-domain, and the other one is defined in the T-F domain.
Our experiment demonstrates the effectiveness of the proposed system comparing to T-F masking and mask-based beamforming.
\end{abstract}

\begin{keywords}
Multi-channel Wiener filtering, Spectrogram consistency, deep neural networks (DNNs)
\end{keywords}

\section{Introduction}
\label{sec:intro}

Speech enhancement has been studied extensively because of its various applications including mobile communication \cite{Loizou2013} and hearing aids \cite{Doclo2010}.
When multiple microphones are available, multi-channel speech enhancement is an effective approach because it takes advantage of spatial information \cite{Gannot2017}.
Recently, deep neural network (DNN)--based multi-channel speech enhancement has gained increasing attention \cite{Sivasankaran2015,Heymann2016,Erdogan2016} motivated by its strong modeling capability.
DNN-based multi-channel speech enhancement methods often manipulate an observed signal in the time-frequency (T-F) domain because spatial filtering can be efficiently implemented in the T-F domain.
Ordinarily, the estimated spectrogram or T-F mask are passed to objective functions defined in the T-F domain.
However, the enhanced time-domain signal is important for human listeners.
Hence, this paper proposes a DNN-based multi-channel speech enhancement system in which speech enhancement is conducted in the T-F domain, and objective functions are computed on the reconstructed time-domain signal to improve human perception.

Recently, various DNN-based approaches to multi-channel speech enhancement have been studied \cite{Sivasankaran2015,Heymann2016,Xiao2016}.
A well-known approach is mask-based beamforming (MB) in which T-F mask is used for estimating the spatial covariance matrix (SCM) \cite{Heymann2016,Erdogan2016}.
Although it achieved excellent performance as the front-end of ASR \cite{Higuchi2017,Watanabe2017,Yoshioka2018}, it has a few drawbacks.
First, the DNN was often trained to minimize the estimation error of T-F masks instead of maximizing the quality of the estimated signal directly \cite{Ochiai2017,Heymann2017}.
In addition, its performance is limited under noisy and reverberant environment because it does not consider non-stationary characteristics of speech signal \cite{ZQWang2018b}.

To address these problems, we proposed a DNN-based multi-channel Wiener filtering (MWF) with a multi-channel objective function for speech separation \cite{Togami2019b}.
The DNN-based MWF is based on the estimation of time-varying SCMs, and it can adapt to the time-varying speech signal.
In addition, the quality of the estimated signal is directly maximized in the T-F domain based on a statistical model of a multi-channel signal.
Hence, it can be expected that the DNN-based MWF improves the performance of multi-channel speech enhancement as in speech separation.
However, its result is often inconsistent \cite{Griffin1984,Roux2008,Masuyama2019a}, and thus the estimated amplitude and phase may change by applying the inverse STFT (iSTFT) and STFT.
Although several DNN-based monaural speech enhancement and separation methods improve the performance by considering consistency \cite{ZQWang2018c,Wisdom2019}, the consistency was not taken into account in DNN-based multi-channel speech enhancement.

\begin{figure}[t]
\centering
\includegraphics[width=0.99\columnwidth]{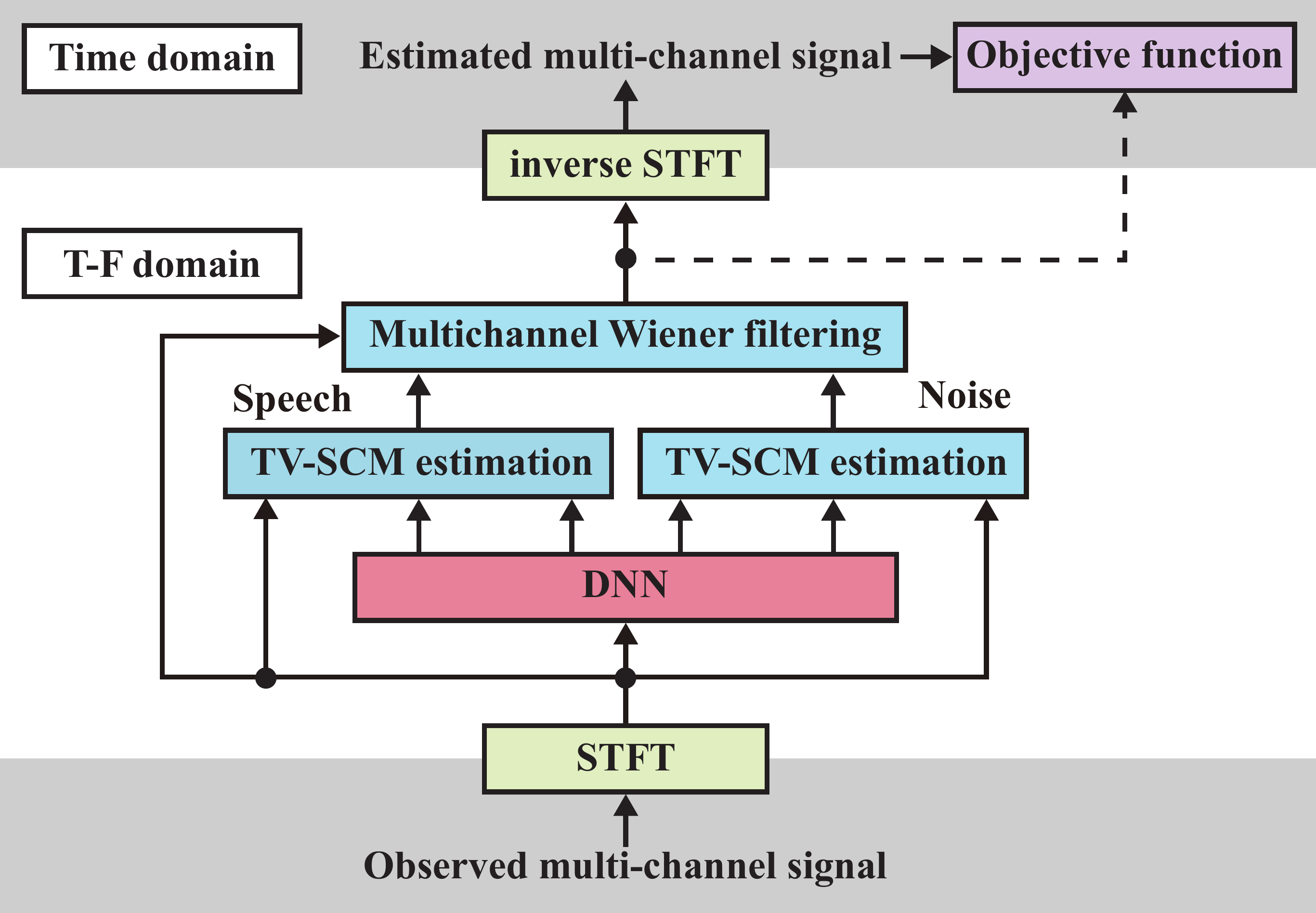}
\caption{Block diagram of proposed multi-channel speech enhancement system. Green and blue blocks indicate STFT-related layers and multi-channel signal processing layers, respectively. Red block represents DNN, and only this block is trainable. DNN is trained to maximize the quality of estimated time-domain signal.}
\label{fig:illust}
\end{figure}

In this paper, we propose a novel system for a DNN-based multi-channel speech enhancement where the DNN is trained to improve the quality of the enhanced time-domain signal directly.
The overview of the proposed system is illustrated in Fig.~\ref{fig:illust}.
The DNN estimates T-F masks and power spectral densities of speech and noise to calculate MWF.
Multi-channel speech enhancement is conducted by MWF, which is represented by blue blocks in Fig.~\ref{fig:illust}.
The estimated spectrogram is converted back to the time-domain and passed to an objective function.
Thanks to this, the objective function can consider the reconstruction error due to the inconsistency.
We investigate two novel objective functions for evaluating the enhanced time-domain signal in the time or T-F domain.
Our experiment confirmed the performance of the DNN-based MWF was improved by using the proposed objective functions.

\section{Preliminaries}

\subsection{Speech enhancement by multi-channel Wiener filtering}

Since our proposed system uses MWF, this subsection reviews MWF that has been applied to multi-channel speech enhancement \cite{Shimada2019}.
Let a noisy signal be observed by $K$ microphones, and $\mathbf{x}_{t, f} \in \mathbb{C}^K$ be the observed noisy signal in the T-F domain where $t = 0, \ldots, T-1$ and $f = 0, \ldots, F-1$ are the time and frequency indices, respectively.
The observed signal is given by the sum of the clean speech $\mathbf{s}_{t, f} \in \mathbb{C}^K$ and noise $\mathbf{n}_{t, f} \in \mathbb{C}^K$ as
\begin{equation}
\mathbf{x}_{t, f} = \mathbf{s}_{t, f} + \mathbf{n}_{t, f}.
\end{equation}
We assume both speech and noise follow multivariate zero-mean complex Gaussian distributions as in \cite{Duong2010}:
\begin{align}
\mathbf{s}_{t, f} &\sim \mathcal{N}_\mathbb{C} (0, \mathbf{R}_{t, f}^{(s)}), \\
\mathbf{n}_{t, f} &\sim \mathcal{N}_\mathbb{C} (0, \mathbf{R}_{t, f}^{(n)}),
\end{align}
where $\mathbf{R}_{t, f}^{(s)} \in \mathbb{S}_+^{K \times K}$ and $\mathbf{R}_{t, f}^{(n)} \in \mathbb{S}_+^{K \times K}$ are the time-varying SCMs of speech and noise, respectively.
The observed noisy signal also follows a multivariate zero-mean complex Gaussian distribution: $\mathbf{x}_{t, f} \sim \mathcal{N}_\mathbb{C} (0, \mathbf{R}_{t, f}^{(s)} + \mathbf{R}_{t, f}^{(n)})$.

Given time-varying SCMs, the posterior distribution $\mathbf{s}_{t, f}|\mathbf{x}_{t, f}$ follows a multivariate complex Gaussian distribution:
\begin{equation}
\mathbf{s}_{t, f}|\mathbf{x}_{t, f}  \sim \mathcal{N}_\mathbb{C} (\boldsymbol{\mu}_{t, f}, \boldsymbol{\Psi}_{t,f}),
\end{equation}
where its mean $\boldsymbol{\mu}_{t, f}$ and covariance matrix $\boldsymbol{\Psi}_{t,f}$ are calculated as
\begin{align}
\boldsymbol{\mu}_{t, f} &= \mathbf{W}_{t,f} \mathbf{x}_{t, f}, \label{eq:mwf} \\
\mathbf{W}_{t,f} &= \mathbf{R}_{t, f}^{(s)}\bigl(\mathbf{R}_{t, f}^{(s)} + \mathbf{R}_{t, f}^{(n)} \bigr)^{-1}, \label{eq:mwfcalc} \\
\boldsymbol{\Psi}_{t,f} &= (\mathbf{I} - \mathbf{W}_{t,f}) \mathbf{R}_{t, f}^{(s)}, \label{eq:mwfcov}
\end{align}
$\mathbf{I} \in \mathbb{R}^{K \times K}$ is the identity matrix, and $\mathbf{W}_{t,f}$ is called MWF.
The enhanced spectrogram is obtained by applying a MWF, and the result is converted back to the time-domain by applying iSTFT.

\subsection{DNN-based multi-channel Wiener filtering}
\label{sec:dnnmwf}

We proposed a DNN-based MWF for taking advantage of the strong modeling capability of a DNN in multi-channel speech separation \cite{Togami2019b}.
In the DNN-based MWF, a DNN estimates T-F mask $\mathbf{M}^{(s_j)} \in \mathbb{R}_+^{T \times F}$ and power spectral density $\mathbf{V}^{(s_j)} \in \mathbb{R}_+^{T \times F}$ for each speaker.
The time-varying SCM of $j$th speaker is calculated by 
\begin{align}
\hat{\mathbf{R}}_{t, f}^{(s_j)} &= V_{t, f}^{(s_j)} \tilde{\mathbf{R}}_{f}^{(s_j)}, \\
\tilde{\mathbf{R}}_{f}^{(s_j)} &= \frac{1}{\sum_{t} M_{t,f}^{(s_j)}} \sum_{t} M_{t,f}^{(s_j)} \mathbf{x}_{t, f} \mathbf{x}_{t,f}^{H},
\end{align}
where $\tilde{\mathbf{R}}_{f}^{(s_j)}$ is a time-invariant SCM estimated by using T-F mask, and $\mathbf{x}_{t,f}^{H}$ is the Hermitian transpose of $\mathbf{x}_{t,f}$.
Based on the estimated time-varying SCMs, each speech signal is estimated by MWF.

To train the DNN for estimating T-F masks and power spectral densities, we proposed the following objective function \cite{Togami2019b}:
\begin{align}
\mathcal{L}_\text{base} &= \sum_{t,f}
\mathbf{d}_{t,f}^{(s_j) H} \hat{\boldsymbol{\Psi}}_{t,f}^{(s_j)-1} \mathbf{d}_{t,f}^{(s_j)} + \log \text{det}(\hat{\boldsymbol{\Psi}}^{(s_j)}_{t,f}) \label{eq:mis}, \\
\mathbf{d}_{t,f}^{(s_j)} &= \mathbf{s}_{t,f}^{(s_j)} - \hat{\mathbf{s}}_{t, f}^{(s_j)},
\end{align}
where $\hat{\mathbf{s}}_{t, f}^{(s_j)}$ is the multi-channel signal estimated by MWF, and $\hat{\boldsymbol{\Psi}}^{(s_j)}_{t,f}$ is the covariance calculated by Eq.~\eqref{eq:mwfcov}.
The minimization of this objective function corresponds to the maximization of the posterior distribution $\mathbf{s}_{t, f}|\mathbf{x}_{t, f}$.
In other words, the objective function given in Eq.~\eqref{eq:mis} evaluates the quality of the estimated multi-channel signal based on the statistical model of multi-channel signals.
One undisputed advantage of this objective function is that the separated signal is directly evaluated while conventional methods have set auxiliary targets, such as T-F mask, in their objective functions \cite{Sivasankaran2015,Heymann2016}.
The effectiveness of the multi-channel objective function has also been confirmed in various MB \cite{Masuyama2019b}.

\subsection{STFT consistency}
\label{sec:consistency}
It is known that spectrograms calculated by STFT have a relation between neighborhood T-F bins, and they are called consistent spectrograms \cite{Griffin1984,Roux2008,Masuyama2019a}.
A consistent spectrogram satisfies the following relation:
\begin{equation}
\mathbf{X} = \mathcal{P}(\mathbf{X}) = \mathcal{G} \circ　\mathcal{G}^\dagger (\mathbf{X}),
\end{equation}
where $\mathcal{G}$ is STFT, $\mathcal{G}^\dagger$ is iSTFT, and $\mathcal{P}(\mathbf{X})$ is the projection onto the set of consistent spectrograms.
When speech enhancement is conducted in the T-F domain, the consistency of the estimated spectrogram is not guaranteed.
In such a case, the spectrogram calculated by STFT of the reconstructed time-domain signal differs from the estimated spectrogram.
In DNN-based speech enhancement, this discrepancy indicates that objective functions defined in the T-F domain do not evaluate the estimated time-domain speech properly.
Some studies have addressed this problem since the discrepancy decreases the performance of T-F masking \cite{ZQWang2018c,Wisdom2019}.
For instance, \cite{ZQWang2018c} presented the wave approximation (WA) which evaluates the estimated signal in the time-domain, and \cite{Wisdom2019} proposed to evaluate a spectrogram projected onto the set of consistent spectrograms.
Although these studies showed the importance of the consistency in monaural speech enhancement and separation, it was not explicitly considered in multi-channel speech enhancement.

\section{Proposed multi-channel speech enhancement system}

\begin{figure}[t!]
\centering
\vspace{2pt}
\includegraphics[width=0.99\columnwidth]{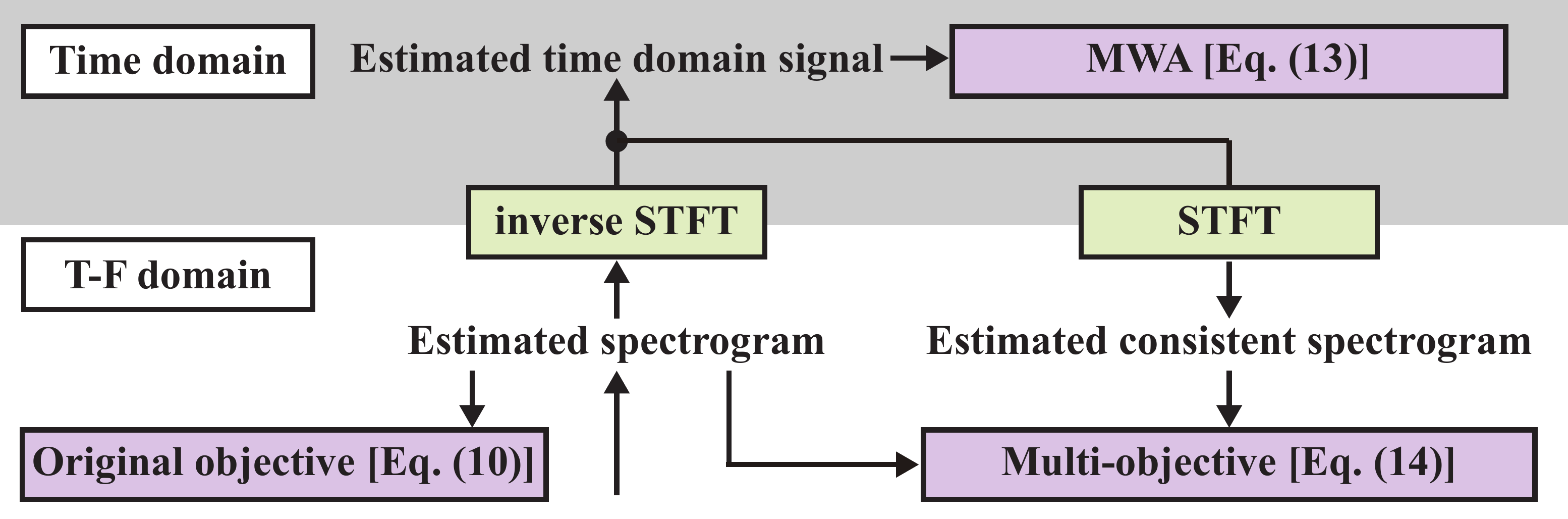}
\vspace{2pt}
\caption{Illustration of proposed objective functions.}
\label{fig:objective}
\end{figure}

In this section, we propose a system of DNN-based multi-channel speech enhancement in which an objective function is computed on the estimated time-domain signal as illustrated in Fig.~\ref{fig:illust}.
In the proposed system, multi-channel speech enhancement is conducted by the DNN-based MWF as described in Section~\ref{sec:dnnmwf}.
Then, the result of MWF is converted back to the time-domain by iSTFT and passed to the objective function.
In Section~\ref{sec:mwa}, we extend WA for applying it to our proposed system, which is defined in the time-domain.
Section~\ref{sec:mof} describes another objective function calculated by a sum of the original multi-channel objective function \cite{Togami2019b} and a consistency-aware objective function defined in the T-F domain.
Both proposed objective functions are summarized in Fig.~\ref{fig:objective}.

\subsection{Multi-channel wave approximation (MWA)}
\label{sec:mwa}

The multi-channel objective function given in Eq.~\eqref{eq:mis} is computed on the estimated spectrogram $\hat{\mathbf{s}}$ in the T-F domain, and it does not consider the reconstruction error due to the inconsistency of the estimated spectrogram.
To address this problem, we propose multi-channel wave approximation (MWA) which is computed on the reconstructed time domain signal as illustrated in Fig.~\ref{fig:objective}.
It is formulated as a sum of WA at each channel:
\begin{equation}
\mathcal{L}_\text{MWA} = \sum_{k} \| \mathcal{G}^\dagger (\mathbf{S}_{k}) - \mathcal{G}^\dagger (\hat{\mathbf{S}}_{k})  \|_1, \label{eq:mwa}
\vspace{-2pt}
\end{equation}
where $\| \cdot \|_1$ is the $\ell_1$ norm, and $\mathbf{S}_{k} \in \mathbb{C}^{T \times F}$ and $\hat{\mathbf{S}}_{k} \in \mathbb{C}^{T \times F}$ are the clean and estimated spectrograms at $k$th channel, respectively.
WMA trains to maximize the quality of the reconstructed time-domain signal while the original objective function given by Eq.~\eqref{eq:mis} focuses on the estimated spectrogram which may be inconsistent.
Recently, WA have achieved promising results in monaural speech enhancement and separation \cite{ZQWang2018c}.
The proposed MWA  is a simple extension of WA to multi-channel case.

\subsection{Consistency-aware multi-channel objective function}
\label{sec:mof}

We propose another consistency-aware multi-channel objective function as a sum of the original multi-channel objective function given in Eq.~\eqref{eq:mis} and a consistency-aware objective function:
\begin{equation}
\mathcal{L}_\text{multi} = \mathcal{L}_\text{base} + \lambda \sum_{k} \| \mathbf{S}_{k} - \mathcal{P}(\hat{\mathbf{S}}_{k}) \|_\mathrm{Fro}^2, \label{eq:multi}
\vspace{-2pt}
\end{equation}
where $\| \cdot \|_\mathrm{Fro}$ is the Frobenius norm, and $\lambda \in \mathbb{R}_+$ is a hyperparameter for adjusting two terms. 
In the second term, the estimated spectrogram is projected onto the set of consistent spectrograms, and then the distance to the clean spectrogram is calculated.
This projection enables the objective functioin to consider the reconstruction error due to the inconsistency.
In other words, the second term corresponds to evaluating the estimated time-domain signal in the T-F domain by recomputing STFT.
Note that the second term in Eq.~\eqref{eq:multi} does not have any known statistical meaning while the first term is based on a statistical model of multi-channel signals.
It can be considered to evaluate the posterior distribution with the consistency projection, which is included in our future work.

The proposed objective functions are summarized in Fig.~\ref{fig:objective}.
The first proposed objective function given in Eq.~\eqref{eq:mwa} is defined between the clean and estimated time-domain signal.
In contrast, the second one given in Eq.~\eqref{eq:multi} considers both estimated spectrogram and STFT of the reconstructed time-domain signal.
In our early experiment, this combination achieved better performance comparing to only using the second term.
The difference of the domain of the proposed objective functions affects the enhanced signal as shown in the following experiment.

\section{Experiments and results}

To confirm the effectiveness of the proposed system, an experiment of multi-channel speech enhancement under diffuse noise was conducted.
DNN-based MWFs were compared with various baseline methods including T-F masking and MB.
In the following subsections, the DNN-based MWFs using DNNs trained by the proposed objective functions given by Eqs.~\eqref{eq:mwa} and \eqref{eq:multi} are refereed to as Prop.~1 and Prop.~2, respectively.

{\renewcommand\arraystretch{1.1}
\begin{table*}[t]
\centering
\footnotesize
\caption{Results of speech enhancement.}
\label{tab:result}
\begin{tabular}{c|cc||ccc||ccc||ccc}
\hline \hline
\multicolumn{3}{c||}{} & \multicolumn{3}{c||}{$SNR = 0$ dB} & \multicolumn{3}{|c||}{$SNR = 6$ dB} & \multicolumn{3}{c}{$SNR = 12$ dB}\\ 
\hline 
$\mathrm{RT}_{60}$ & \multicolumn{2}{c||}{Approach} & SDR [dB] & CD [dB] & PESQ & SDR [dB] & CD [dB] &PESQ & SDR [dB] & CD [dB] & PESQ \\
\hline \hline
\multirow{8}{*}{$160$ ms}
& 
& Observed & 1.14 & 5.26 & 1.14 &6.93 & 4.67 & 1.33 & 13.21 & 3.74 & 1.77 \\
\cline{2-12}
& \multirow{3}{*}{T-F masking}
& PSA & 9.15 & 4.42 & 1.61 & 13.46 & 3.70 & 2.00 & 18.07 & 2.95 & 2.59 \\
&
& PSA+Proj & 9.36 & 4.53 & 1.62 & 13.76 & 3.80 & 2.03 & 18.40 & 3.08 & 2.64 \\
&
& WA & 9.60 & 4.65 & 1.63 & 14.06 & 3.90 & 2.11 & 18.69 & 3.15 & 2.73 \\
\cline{2-12}
& \multirow{4}{*}{Spatial filtering}
& MB & 5.42 & 4.88 & 1.28 & 11.54 & 4.22 & 1.65 & 16.85 & 3.24 & 2.27 \\
\cline{4-12}
&
& Original & 10.48 & 4.41 & 1.77 & 14.90 & 3.62 & 2.26 & 19.20 & 2.71 & 2.81 \\
&
& Prop.~1 & \textbf{11.23} & 4.73 & 1.73 & \textbf{15.57} & 3.97 & 2.23 & \textbf{19.75} & 3.15 & 2.84 \\
&
& Prop.~2 & 10.72 & \textbf{4.38} & \textbf{1.82} & 15.01 & \textbf{3.60} & \textbf{2.30} & 19.39 & \textbf{2.68} & \textbf{2.85} \\
\hline \hline
\multirow{8}{*}{$360$ ms}
&
& Observed & 0.84 & 5.28 & 1.12 & 6.77 & 4.58 & 1.33 & 12.87 & 3.77 & 1.75 \\
\cline{2-12}
& \multirow{3}{*}{T-F masking}
& PSA & 9.11 & 4.45 & 1.60 & 13.38 & 3.65 & 2.03 & 17.90 & 2.96 & 2.57 \\
&
& PSA+Proj & 9.27 & 4.53 & 1.60 & 13.66 & 3.75 & 2.05 & 18.22 & 3.06 & 2.62 \\
&
& WA & 9.56 & 4.65 & 1.62 & 13.99 & 3.85 & 2.15 & 18.54 & 3.14 & 2.75 \\
\cline{2-12}
& \multirow{4}{*}{Spatial filtering}
& MB & 5.20 & 4.90 & 1.26 & 11.07 & 4.15 & 1.62 & 16.34 & 3.32 & 2.19 \\
\cline{4-12}
&
& Original & 10.38 & 4.42 & 1.76 & 14.50 & 3.58 & 2.23 & 18.82 & 2.76 & 2.77 \\
&
& Prop.~1 & \textbf{11.23} & 4.73 & 1.71 & \textbf{15.32} & 3.92 & 2.25 & \textbf{19.56} & 3.19 & 2.79 \\
&
& Prop.~2 & 10.67 & \textbf{4.40} & \textbf{1.80} & 14.71 & \textbf{3.57} & \textbf{2.29} & 18.98 & \textbf{2.74} & \textbf{2.80} \\
\hline \hline
\end{tabular}
\end{table*}
}

\subsection{Experimental setup}

\begin{figure}[t!]
\centering
\includegraphics[width=0.99\columnwidth]{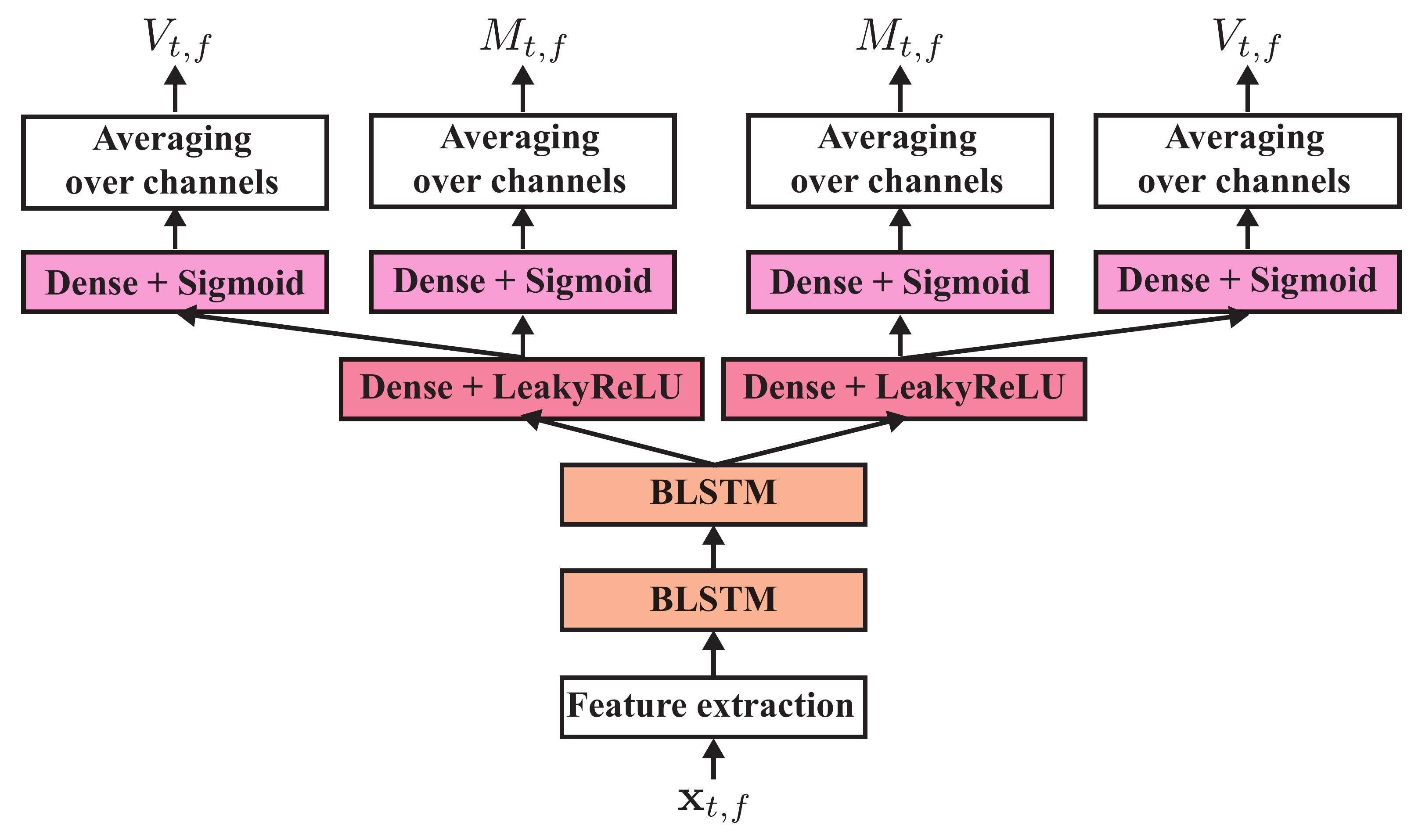}
\caption{Network architecture used in experiment. Only colored blocks contained trainable parameters.
In T-F masking and MB, only the T-F mask of speech $\mathbf{M}^{(s)}$ was used.}
\label{fig:net}
\end{figure}

\subsubsection{Dataset}

In both training and testing, the clean speech in TIMIT corpus \cite{TIMIT} and noise from Diverse Environments Multichannel Acoustic Noise Database (DEMAND) \cite{DEMAND} were used.
The measured impulse responses in Multichannel Impulse Response Database (MIRD) \cite{MIRD} were convoluted to the above dry sources where the $1$st channel of the noise in DEMAND was used as the dry source.
The distance between the speaker and microphones was set to $1$ m, and the azimuth of each talker is randomly selected from $13$ points (from $-90^\circ$ to $90^\circ$ with the intervals of $15^\circ$).
On the other hand, diffuse noise was generated by playing noise from all points.
Note that the noise played at each point is obtained by splitting the original noise into $26$ periods.
The first half was used in the training/validation and the other was used in the testing.
The number of microphones was $2$ where the distance between microphones was  set to $3$ cm.

A training set with $4000$ speech files was randomly selected from the training set of TIMIT, and  the others were used as a validation set.
Since the number of noise was small, we conducted a data augmentation%
\footnote{
The diffuse noise was augmented by conducting convex combinations of two noises, randomly selected from DEMAND, as $\mathbf{n}_{t,f} = \alpha * \mathbf{n}_{t,f}^{(0)} + (1-\alpha) * \mathbf{n}_{t,f}^{(1)}$, where $0 \leq \alpha \leq 1$ is randomly generated from a Beta distribution.
}.
The signal-to-noise ratio (SNR) of the training/validation set was adjusted from $-6$ to $12$ dB.
At the training, the reverberation time ($\text{RT}_{60}$) was $160$ ms.
On the other hand, at the testing, $300$ speeches randomly selected from the testing set of TIMIT were used as clean speach, and the later $13$ periods of the noise were used.
We evaluated under two reverberation conditions: $\text{RT}_{60} = 160$ ms and $\text{RT}_{60} = 360$ ms.
All the speeches were sampled at $16$ kHz, and STFT was computed using the Hann window whose length was $32$ ms with $8$ ms shift.

\subsubsection{Baseline methods}

We compared the proposed methods with the following baseline methods.
At first, T-F masking was used as a well-known monaural speech enhancement approach.
To confirm the effectiveness of considering the consistency, three objective functions [the phase sensitive approximation (PSA) \cite{Erdogan2015}, PSA with the consistency projection (PSA+Proj) \cite{Wisdom2019}, and WA \cite{ZQWang2018c}] were compared.
MB \cite{Heymann2016} was also conducted which used a DNN trained based on PSA.
Although several iterative methods using DNN have been proposed in multi-channel source separation \cite{Sivasankaran2015,Nugraha2016,Makishima2019}, we only compared the proposed system with aforementioned non-iterative methods because it is non-iterative.
The performance of the proposed method can be improved by unifying iterative methods.

\subsubsection{DNN architecture and setup}
In all methods, including T-F masking, the input feature was the concatenation of the amplitude feature and phase-difference features.
The amplitude feature was calculated by
\begin{equation}
\Phi_{k} = \left[\, \mathcal{U}(\log_{10} ( |\mathbf{X}_k| + \delta) \right],
\end{equation}
where $\delta = 0.0001$, and $\mathcal{U}$ is the utterance-level mean and variance normalization.
As in a previous study \cite{ZQwang2018a}, the phase-difference between two microphones was also used as a input feature:
\begin{align}
\cos\mathrm{IPD}_{t, f} &= \cos \left( \mathrm{Arg}(x_{t, f, 1}) - \mathrm{Arg}(x_{t, f, 0}) \right), \\
\sin\mathrm{IPD}_{t, f} &= \sin \left( \mathrm{Arg}(x_{t, f, 1}) - \mathrm{Arg}(x_{t, f, 0}) \right),
\end{align}
where $\mathrm{Arg}(\cdot)$ is the complex argument.

The DNN for the proposed methods is illustrated in Fig.~\ref{fig:net}, which contains two bidirectional long-short term memory (BLSTM) layers and dense layers.
Dropout of $0.3$ was applied to each BLSTM layer and dense layer without the last layers.
The networks are trained on $128$-frame segments using the Adam optimizer over $200$ epochs.
The learning rate was decayed by multiplying $0.5$ if the objective function on the validation set did not decrease for $3$ consecutive epochs, and the initial learning rate was set to $0.0001$.
In Prop. 2, $\lambda$ was set to $1$.
In baseline methods, we used only the T-F mask estimation part of the DNN illustrated in Fig.~\ref{fig:net}.

Note that all systems were implemented using TensorFlow in which STFT and iSTFT are implemented with their backpropagation.
In addition, it  supports a lot of complex-valued operations and their derivatives.
Hence, we can easily apply MWF in the training.

\subsection{Experimental results}
The performances of multi-channel speech enhancement were evaluated by the signal-to-distortion ratio (SDR), cepstrum distortion (CD), and PESQ.
The experimental results are summarized in Table~\ref{tab:result} in which the bold font represents the best score in each condition.
As can be seen from both reverberation conditions, MB resulted in the lowest performance because it does not consider non-stationary characteristics of speech.
In T-F masking, consistency-aware methods, PSA+Proj and WA, outperformed the original PSA in terms of SDR and PESQ.
This results confirmed the importance of the consistency.

The DNN-based MWF with the original multi-channel objective function (Original) \cite{Togami2019b} outperformed the other conventional methods.
Furthermore, the DNN-based MWF with the proposed MWA, Prop.~1, significantly improved SDR. 
On the other hand, by using the multi-objective function given in Eq.~\eqref{eq:multi}, Prop.~2 outperformed the original DNN-based MWF in terms of not only SDR but also CD and PESQ.
We stress that, the difference between three DNN-based MWFs is only the objective function, and thus the computational cost for the inference is the same.

\section{Conclusion}
In this paper, we described the system of DNN-based multi-channel speech enhancement where the DNN is trained to maximize the quality of the time-domain signal estimated by the DNN-based MWF.
We further proposed two objective functions defined on the enhanced time-domain signal.
Our experimental results confirmed the effectiveness of the DNN-based MWF and proposed objective functions in multi-channel speech enhancement.
Future work includes combining the proposed system with iterative algorithms.

\bibliographystyle{IEEEbib}

\begin{thebibliography}{10}

\bibitem{Loizou2013}
P.~C. Loizou,
\newblock {\em Speech Enhancement: Theory and Practice, Second Edition},
\newblock CRC Press, Inc., 2nd edition, Feb. 2013.

\bibitem{Doclo2010}
S.~Doclo, S.~Gannot, M.~Moonen, and A.~Spriet,
\newblock {\em Handbook on Array Processing and Sensor Network}, chapter
  Acoustic Beamforming for Hearing Aid Applications, pp. 269--302,
\newblock Wiley Online Library, Jan. 2010.

\bibitem{Gannot2017}
S.~Gannot, E.~Vincent, S.~Markovich-Golan, and A.~Ozerov,
\newblock ``A consolidated perspective on multimicrophone speech enhancement
  and source separation,''
\newblock {\em IEEE/ACM Trans. Audio, Speech and Lang. Proc.}, vol. 25, no. 4,
  pp. 692--730, Apr. 2017.

\bibitem{Sivasankaran2015}
S.~{Sivasankaran}, A.~A. {Nugraha}, E.~{Vincent}, J.~A. {Morales-Cordovilla},
  S.~{Dalmia}, I.~{Illina}, and A.~{Liutkus},
\newblock ``Robust {A}{S}{R} using neural network based speech enhancement and
  feature simulation,''
\newblock in {\em IEEE Workshop Autom. Speech Recognit. Underst. (ASRU)}, Dec.
  2015, pp. 482--489.

\bibitem{Heymann2016}
J.~{Heymann}, L.~{Drude}, and R.~{Haeb-Umbach},
\newblock ``Neural network based spectral mask estimation for acoustic
  beamforming,''
\newblock in {\em IEEE Int. Conf. on Acoust., Speech Signal Process. (ICASSP)},
  Mar. 2016, pp. 196--200.

\bibitem{Erdogan2016}
H.~{Erdogan}, J.~R. {Hershey}, S.~{Watanabe}, M.~Mandel, and J.~{Le Roux},
\newblock ``Improved mvdr beamforming using single-channel mask prediction
  networks,''
\newblock in {\em INTERSPEECH}, Sept. 2016, pp. 1981--1985.

\bibitem{Xiao2016}
X.~{Xiao}, S.~{Watanabe}, H.~{Erdogan}, L.~{Lu}, J.~{Hershey}, M.~L. {Seltzer},
  G.~{Chen}, Y.~{Zhang}, M.~{Mandel}, and D.~{Yu},
\newblock ``Deep beamforming networks for multi-channel speech recognition,''
\newblock in {\em IEEE Int. Conf. on Acoust., Speech Signal Process. (ICASSP)},
  2016, pp. 5745--5749.

\bibitem{Higuchi2017}
T.~{Higuchi}, N.~{Ito}, S.~{Araki}, T.~{Yoshioka}, M.~{Delcroix}, and
  T.~{Nakatani},
\newblock ``Online {M}{V}{D}{R} beamformer based on complex {G}aussian mixture
  model with spatial prior for noise robust {A}{S}{R},''
\newblock {\em IEEE/ACM Trans. Audio, Speech, Language Process.}, vol. 25, no.
  4, pp. 780--793, Apr. 2017.

\bibitem{Watanabe2017}
S.~Watanabe, M.~Delcroix, F.~Metze, and J.~R. Hershey,
\newblock {\em New Era for Robust Speech Recognition: {E}xploiting Deep
  Learning},
\newblock Springer, 2017.

\bibitem{Yoshioka2018}
T.~{Yoshioka}, H.~{Erdogan}, Z.~{Chen}, and F.~{Alleva},
\newblock ``Multi-microphone neural speech separation for far-field
  multi-talker speech recognition,''
\newblock in {\em IEEE Int. Conf. on Acoust., Speech Signal Process. (ICASSP)},
  Apr. 2018, pp. 5739--5743.

\bibitem{Ochiai2017}
Tsubasa Ochiai, Shinji Watanabe, Takaaki Hori, and John~R. Hershey,
\newblock ``Multichannel end-to-end speech recognition,''
\newblock in {\em Int. Conf. Mach. Learn. (ICML)}, Aug. 2017, pp. 2632--2641.

\bibitem{Heymann2017}
J.~{Heymann}, L.~{Drude}, C.~{Boeddeker}, P.~{Hanebrink}, and R.~{Haeb-Umbach},
\newblock ``Beamnet: End-to-end training of a beamformer-supported
  multi-channel {A}{S}{R} system,''
\newblock in {\em IEEE Int. Conf. on Acoust., Speech Signal Process. (ICASSP)},
  2017, pp. 5325--5329.

\bibitem{ZQWang2018b}
Z.~Wang and D.~Wang,
\newblock ``All-neural multi-channel speech enhancement,''
\newblock in {\em Interspeech}, Sept. 2018, pp. 3234--3238.

\bibitem{Togami2019b}
M.~{Togami},
\newblock ``Multi-channel {I}takura {S}aito distance minimization with deep
  neural network,''
\newblock in {\em IEEE Int. Conf. on Acoust., Speech Signal Process. (ICASSP)},
  May 2019, pp. 536--540.

\bibitem{Griffin1984}
D.~Griffin and J.~Lim,
\newblock ``Signal estimation from modified short-time {F}ourier transform,''
\newblock {\em IEEE Trans. Acoust., Speech, Signal Process.}, vol. 32, no. 2,
  pp. 236--243, Apr. 1984.

\bibitem{Roux2008}
J.~{{Le} Roux}, N.~Ono, and S.~Sagayama,
\newblock ``Explicit consistency constraints for {S}{T}{F}{T} spectrograms and
  their application to phase reconstruction,''
\newblock in {\em ISCA Workshop Stat. Percept. Audit. (SAPA)}, Sept. 2008, pp.
  23--28.

\bibitem{Masuyama2019a}
Y.~Masuyama, K.~Yatabe, and Y.~Oikawa,
\newblock ``{G}riffin--{L}im like phase recovery via alternating direction
  method of multipliers,''
\newblock {\em IEEE Signal Process. Lett.}, vol. 26, no. 1, pp. 184--188, Jan.
  2019.

\bibitem{ZQWang2018c}
Z.~Wang, D.~Wang J.~{Le Roux}, and J.~Hershey,
\newblock ``End-to-end speech separation with unfolded iterative phase
  reconstruction,''
\newblock in {\em Interspeech}, Sept. 2018, pp. 2708--2712.

\bibitem{Wisdom2019}
S.~{Wisdom}, J.~R. {Hershey}, K.~{Wilson}, J.~{Thorpe}, M.~{Chinen},
  B.~{Patton}, and R.~A. {Saurous},
\newblock ``Differentiable consistency constraints for improved deep speech
  enhancement,''
\newblock in {\em IEEE Int. Conf. on Acoust., Speech Signal Process. (ICASSP)},
  May 2019, pp. 900--904.

\bibitem{Shimada2019}
K.~{Shimada}, Y.~{Bando}, M.~{Mimura}, K.~{Itoyama}, K.~{Yoshii}, and
  T.~{Kawahara},
\newblock ``Unsupervised speech enhancement based on multichannel
  {N}{M}{F}-informed beamforming for noise-robust automatic speech
  recognition,''
\newblock {\em IEEE/ACM Trans. Audio, Speech, Lang. Process.}, vol. 27, no. 5,
  pp. 960--971, May 2019.

\bibitem{Duong2010}
N.~Q.~K. {Duong}, E.~{Vincent}, and R.~{Gribonval},
\newblock ``Under-determined reverberant audio source separation using a
  full-rank spatial covariance model,''
\newblock {\em IEEE Trans Audio, Speech, Lang. Process.}, vol. 18, no. 7, pp.
  1830--1840, Sept. 2010.

\bibitem{Masuyama2019b}
Y.~Masuyama, M.~Togami, and T.~Komatsu,
\newblock ``Multichannel loss function for supervised speech source separation
  by mask-based beamforming,''
\newblock in {\em Interspeech}, Sept. 2019.

\bibitem{TIMIT}
J.~S. Garofolo, L.~F. Lamel, W.~M. Fisher, J.~G Fiscus, and D.~S. Pallett,
\newblock ``{D}{A}{R}{P}{A} {T}{I}{M}{I}{T} acoustic-phonetic continous speech
  corpus {C}{D}-{R}{O}{M},''
\newblock 1993.

\bibitem{DEMAND}
J.~Thiemann, N.~Ito, and E.~Vincent,
\newblock ``The diverse environments multi-channel acoustic noise database: A
  database of multichannel environmental noise recording,''
\newblock {\em J. Acoust. Soc. Am.}, vol. 133, no. 5, pp. 3591--3591, 2013.

\bibitem{MIRD}
E.~{Hadad}, F.~{Heese}, P.~{Vary}, and S.~{Gannot},
\newblock ``Multichannel audio database in various acoustic environments,''
\newblock in {\em Int. Workshop Acoust. Signal Enhance. (IWAENC)}, Sept. 2014,
  pp. 31--317.

\bibitem{Erdogan2015}
H.~{Erdogan}, J.~R. {Hershey}, S.~{Watanabe}, and J.~{Le Roux},
\newblock ``Phase-sensitive and recognition-boosted speech separation using
  deep recurrent neural networks,''
\newblock in {\em IEEE Int. Conf. on Acoust., Speech Signal Process. (ICASSP)},
  Apr. 2015, pp. 708--712.

\bibitem{Nugraha2016}
A.~A. {Nugraha}, A.~{Liutkus}, and E.~{Vincent},
\newblock ``Multichannel audio source separation with deep neural networks,''
\newblock in {\em IEEE Int. Conf. on Acoust., Speech Signal Process. (ICASSP)},
  Sept. 2016, vol.~24, pp. 1652--1664.

\bibitem{Makishima2019}
N.~{Makishima}, S.~{Mogami}, N.~{Takamune}, D.~{Kitamura}, H.~{Sumino},
  S.~{Takamichi}, H.~{Saruwatari}, and N.~{Ono},
\newblock ``Independent deeply learned matrix analysis for determined audio
  source separation,''
\newblock {\em IEEE/ACM Trans. Audio, Speech Lang.Process.}, vol. 27, no. 10,
  pp. 1601--1615, Oct. 2019.

\bibitem{ZQwang2018a}
Z.~{Wang}, J.~{Le Roux}, and J.~R. {Hershey},
\newblock ``Multi-channel deep clustering: Discriminative spectral and spatial
  embeddings for speaker-independent speech separation,''
\newblock in {\em IEEE Int. Conf. on Acoust., Speech Signal Process. (ICASSP)},
  Apr. 2018, pp. 1--5.

\end{thebibliography}

\end{document}